\documentclass[runningheads]{llncs}
\usepackage{graphicx}
\usepackage{amssymb}
\usepackage{amsmath}
\usepackage[colorlinks]{hyperref}
\usepackage{tikz}
\usepackage{tabularx}
\usepackage{xcolor}

\begin{document}

\renewcommand{\arraystretch}{1.4}
\setlength{\tabcolsep}{0.3em}

\title{Swin UNETR: Swin Transformers for Semantic Segmentation of Brain Tumors in MRI Images}
\titlerunning{Swin Transformers for Semantic Segmentation of Brain Tumors}
\authorrunning{Hatamizadeh et al.}
\author{Ali Hatamizadeh\inst{1} \and Vishwesh Nath\inst{1} \and Yucheng Tang\inst{2} \and Dong Yang\inst{1} \and \\Holger R. Roth\inst{1}\and Daguang Xu\inst{1}}
\authorrunning{Hatamizadeh et al.}
\institute{NVIDIA
\and
Vanderbilt University\\
\email{ahatamizadeh@nvidia.com} 
}

%
\maketitle
\begin{abstract}

Semantic segmentation of brain tumors is a fundamental medical image analysis task involving multiple MRI imaging modalities that can assist clinicians in diagnosing the patient and successively studying the progression of the malignant entity. In recent years, Fully Convolutional Neural Networks (FCNNs) approaches have become the de facto standard for 3D medical image segmentation. The popular ``U-shaped'' network architecture has achieved state-of-the-art performance benchmarks on different 2D and 3D semantic segmentation tasks and across various imaging modalities. However, due to the limited kernel size of convolution layers in FCNNs, their performance of modeling long-range information is sub-optimal, and this can lead to deficiencies in the segmentation of tumors with variable sizes. On the other hand, transformer models have demonstrated excellent capabilities in capturing such long-range information in multiple domains, including natural language processing and computer vision. Inspired by the success of vision transformers and their variants, we propose a novel segmentation model termed Swin UNEt TRansformers (Swin UNETR). Specifically, the task of 3D brain tumor semantic segmentation is reformulated as a sequence to sequence prediction problem wherein multi-modal input data is projected into a 1D sequence of embedding and used as an input to a hierarchical Swin transformer as the encoder. The swin transformer encoder extracts features at five different resolutions by utilizing shifted windows for computing self-attention and is connected to an FCNN-based decoder at each resolution via skip connections. We have participated in BraTS 2021 segmentation challenge, and our proposed model ranks among the top-performing approaches in the validation phase. \\
Code: \href{https://monai.io/research/swin-unetr}{https://monai.io/research/swin-unetr}

\keywords{Image Segmentation \and Vision Transformer \and Swin Transformer \and UNETR \and Swin UNETR \and BRATS \and Brain Tumor Segmentation}

\end{abstract}

\section{Introduction}
There are over 120 types of brain tumors that affect the human brain~\cite{louis20072007}. As we enter the era of Artificial Intelligence (AI) for healthcare, AI-based intervention for diagnosis and surgical pre-assessment of tumors is at the verge of becoming a necessity rather than a luxury. Elaborate characterization of brain tumors with techniques such as volumetric analysis is useful to study their progression and assist in pre-surgical planning \cite{hoover2011use}. In addition to surgical applications, characterization of delineated tumors can be directly utilized for the prediction of life expectancy~\cite{nie20163d}. Brain tumor segmentation is at the forefront of all such applications.

Brain tumors are categorized into primary and secondary tumor types.  Primary brain tumors originate from brain cells, while secondary tumors metastasize into the brain from other organs. The most common primary brain tumors are gliomas, which arise from brain glial cells and are characterized into low-grade (LGG) and high-grade (HGG) subtypes. High grade gliomas are an aggressive type of malignant brain tumors that grow rapidly and typically require surgery and radiotherapy and have poor survival prognosis~\cite{zacharaki2009classification}. As a reliable diagnostic tool, Magnetic Resonance Imaging (MRI) plays a vital role in monitoring and surgery planning for brain tumor analysis. Typically, several complimentary 3D MRI modalities, such as T1, T1 with contrast agent (T1c), T2 and Fluid-attenuated Inversion Recovery (FLAIR), are required to emphasize different tissue properties and areas of tumor spread.  For instance, gadolinium as the contrast agent emphasizes hyperactive tumor sub-regions in the T1c MRI modality~\cite{grover2015magnetic}.  

Furthermore, automated medical image segmentation techniques~\cite{huo20193d} have shown prominence for providing an accurate and reproducible solution for brain tumor delineation. Recently, deep learning-based brain tumor segmentation techniques~\cite{Myronenko18,jiang2019two,myronenko2019robust,isensee2020nnu} have achieved state-of-the-art performance in various benchmarks~\cite{bakas2018identifying,simpson2019large,baid2021rsna}. These advances are mainly due to the powerful feature extraction capabilities of Convolutional Neural Networks (CNN)s. However, the limited kernel size of CNN-based techniques restricts their capability of learning long-range dependencies that are critical for accurate segmentation of tumors that appear in various shapes and sizes. Although several efforts~\cite{liu2018brain,chen20193d} have tried to address this limitation by increasing the receptive field of the convolutional kernels, the effective receptive field is still limited to local regions.

Recently, transformer-based models have shown prominence in various domains such as natural language processing and computer vision~\cite{vaswani2017attention,devlin2018bert,dosovitskiy2020image}. In computer vision, Vision Transformers~\cite{dosovitskiy2020image}~(ViT)s have demonstrated state-of-the-art performance on various benchmarks. Specifically, self-attention module in ViT-based models allows for modeling long-range information by pairwise interaction between token embeddings and hence leading to more effective local and global contextual representations~\cite{raghu2021vision}. In addition, ViTs have achieved success in effective learning of pretext tasks for self-supervised pre-training in various applications~\cite{caron2021emerging,bao2021beit,tang2021self}. In medical image analysis, UNETR~\cite{hatamizadeh2021unetr} is the first methodology that utilizes a ViT as its encoder without relying on a CNN-based feature extractor. Other approaches~\cite{xie2021cotr,wang2021transbts} have attempted to leverage the power of ViTs as a stand-alone block in their architectures which otherwise consist of CNN-based components. However, UNETR has shown better performance in terms of both accuracy and efficiency in different medical image segmentation tasks~\cite{hatamizadeh2021unetr}.          

Recently, Swin transformers~\cite{liu2021swin,liu2021video} have been proposed as a hierarchical vision transformer that computes self-attention in an efficient shifted window partitioning scheme. As a result, Swin transformers are suitable for various downstream tasks wherein the extracted multi-scale features can be leveraged for further processing. In this work, we propose a novel architecture termed Swin UNEt TRansformers (Swin UNETR), which utilizes a U-shaped network with a Swin transformer as the encoder and connects it to a CNN-based decoder at different resolutions via skip connections. We validate the effectiveness of our approach for the task of multi-modal 3D brain tumor segmentation in the 2021 edition of the Multi-modal Brain Tumor Segmentation Challenge (BraTS). Our model is one of the top-ranking methods in the validation phase and has demonstrated competitive performance in the testing phase. 

\section{Related work}
\label{sec:relatedwork}

In the previous BraTS challenges, ensembles of U-Net shaped architectures have achieved promising results for multi-modal brain tumor segmentation. Kamnitsas~\textit{et al.}~\cite{Kamnitsas17} proposed
a robust segmentation model by aggregating the outputs of various CNN-based models such as 3D U-Net~\cite{cciccek20163d}, 3D FCN~\cite{long2015fully} and Deep Medic~\cite{kamnitsas2017efficient}. Subsequently, Myronenko~\textit{et al.}~\cite{Myronenko18} introduced SegResNet, which utilizes a residual encoder-decoder architecture in which an auxiliary branch is used to reconstruct the input data with a variational auto-encoder as a surrogate task. Zhou~\textit{et al.}~\cite{Zhou18brats} proposed to use an ensemble of different CNN-based networks by taking into account the multi-scale contextual information through an attention block. Zhou~\textit{et al.}~\cite{jiang2019two} used a two-stage cascaded approach consisting of U-Net models wherein the first stage computes a coarse segmentation prediction which will be refined by the second stage. Furthermore, Isensee~\textit{et al.}~\cite{Isensee2020nnUNetFB} proposed the nnU-Net model and demonstrated that a generic U-Net architecture with minor modifications is enough to achieve competitive performance in multiple BraTS challenges.    

Transformer-based models have recently gained a lot of attraction in computer vision~\cite{dosovitskiy2020image,zheng2021rethinking,liu2021swin} and medical image analysis~\cite{chen2021transunet,hatamizadeh2021unetr}. Chen~\textit{et al.}~\cite{chen2021transunet} introduced a 2D U-Net architecture that benefits from a ViT in the bottleneck of the network. Wang~\textit{et al.}~\cite{wang2021transbts} extended this approach for 3D brain tumor segmentation. In addition, Xie~\textit{et al.}~\cite{xie2021cotr} proposed to use a ViT-based model with deformable transformer layers between its CNN-based encoder and decoder by processing the extracted features at different resolutions. Different from these approaches, Hatamizadeh~\textit{et al.}~\cite{hatamizadeh2021unetr} proposed the UNETR architecture in which a ViT-based encoder, which directly utilizes 3D input patches, is connected to a CNN-based decoder. UNETR has shown promising results for brain tumor segmentation using the MSD dataset~\cite{antonelli2021medical}. Unlike the UNETR model, our proposed Swin UNETR architecture uses a Swin transformer encoder which extracts feature representations at several resolutions with a shifted windowing mechanism for computing the self-attention. We demonstrate that Swin transformers~\cite{liu2021swin} have a great capability of learning multi-scale contextual representations and modeling long-range dependencies in comparison to ViT-based approaches with fixed resolution.

\begin{figure*}[t]
\centering
\includegraphics[width=\textwidth]{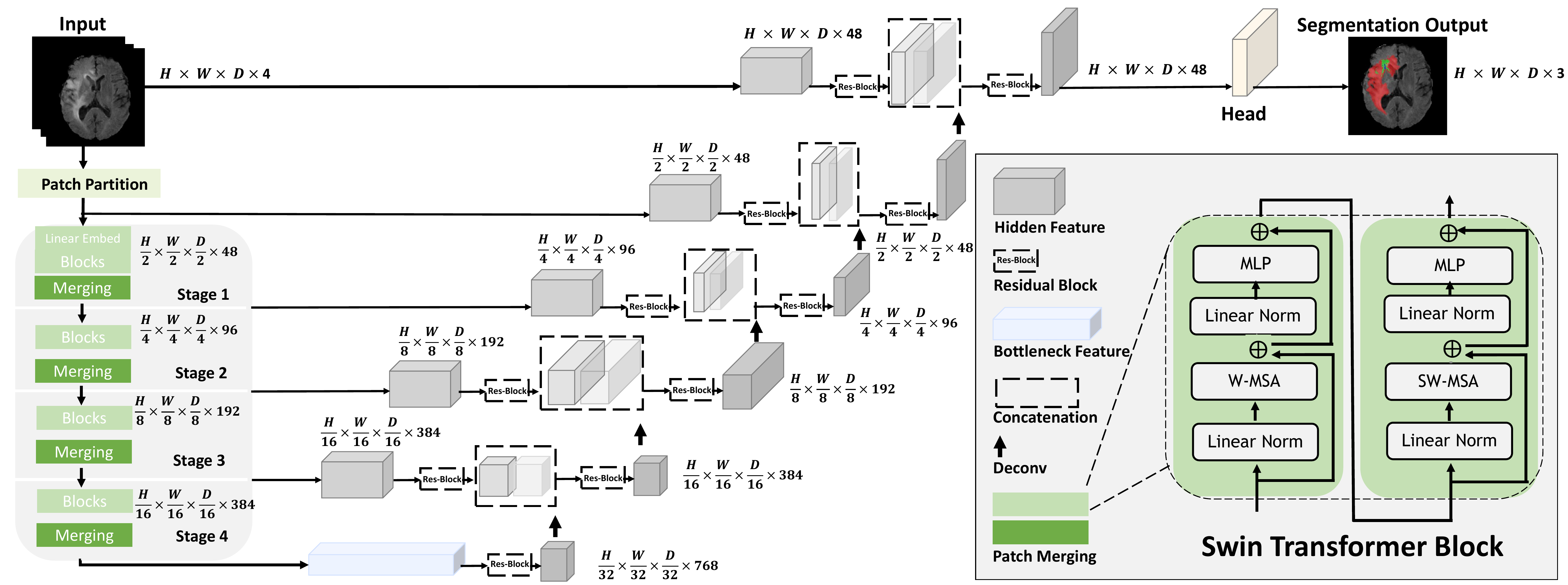}
  \caption{Overview of the Swin UNETR architecture. The input to our model is 3D multi-modal MRI images with 4 channels. The Swin UNETR creates non-overlapping patches of the input data and uses a patch partition layer to create windows with a desired size for computing the self-attention. The encoded feature representations in the Swin transformer are fed to a CNN-decoder via skip connection at multiple resolutions. Final segmentation output consists of 3 output channels corresponding to ET,WT and TC sub-regions.}
  \label{fig:fig2}
\end{figure*}

\section{Swin UNETR}

\label{sec:methods}

\subsection{Encoder}
\begin{figure*}[t]
\centering
\includegraphics[width=\textwidth]{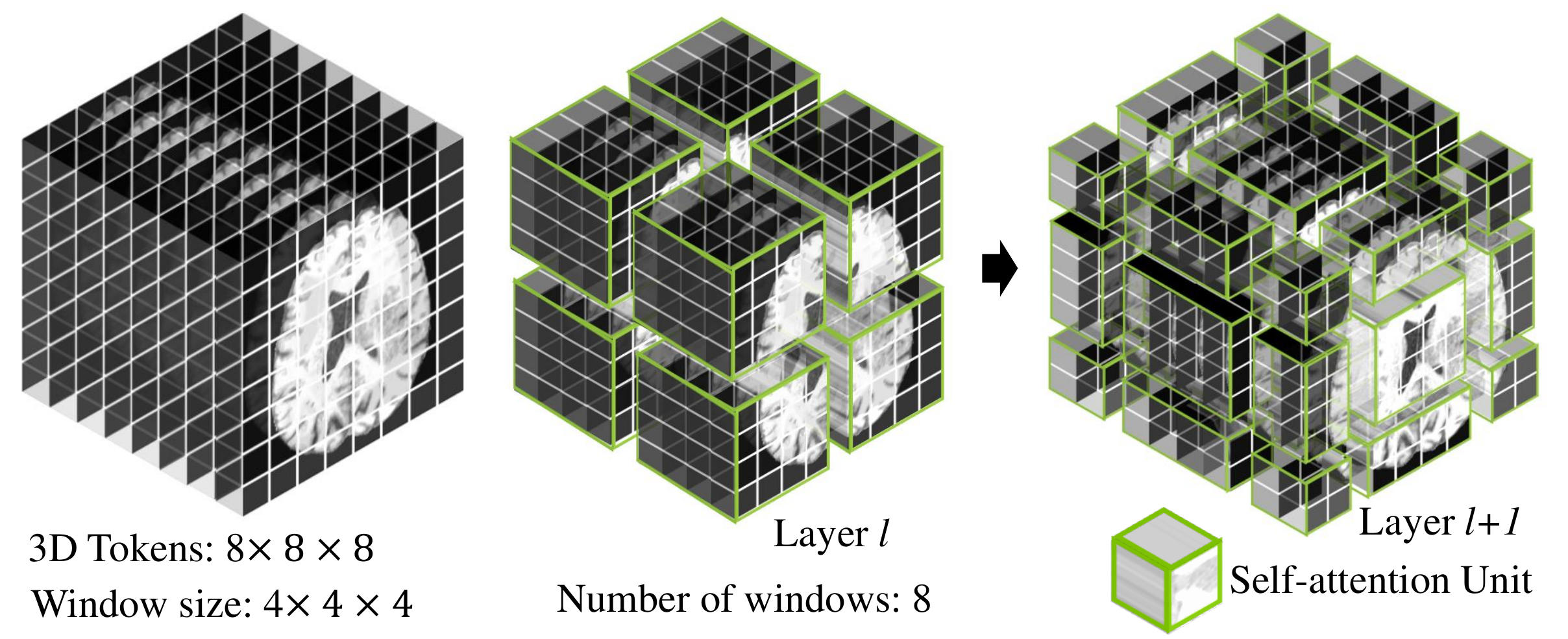}
  \caption{Overview of the shifted windowing mechanism. Note that $8 \times 8 \times 8$ 3D tokens and $4 \times 4 \times 4$ window size are illustrated.}
  \label{fig:fig_attention}
\end{figure*}

We illustrate the architecture of Swin UNETR in Fig.~\ref{fig:fig2}. The input to the Swin UNETR model $\mathcal{X} \in \mathbb{R}^{H\times{W}\times{D}\times{S}}$ is a token with a patch resolution of $(H^{\prime},W^{\prime},D^{\prime})$ and dimension of $H^{\prime} \times W^{\prime}\times D^{\prime}\times S$. We first utilize a patch partition layer to create a sequence of 3D tokens with dimension of $\left\lceil\frac{H}{H^{\prime}}\right\rceil\times\left\lceil\frac{W}{W^{\prime}}\right\rceil\times\left\lceil\frac{D}{D^{\prime}}\right\rceil$ and project them into an embedding space with dimension $C$. The self-attention is computed into non-overlapping windows that are created in the partitioning stage for efficient token interaction modeling. Fig.~\ref{fig:fig_attention} shows the shifted windowing mechanism for subsequent layers. Specifically, we utilize windows of size $M\times M\times M$ to evenly partition a 3D token into $\left\lceil\frac{H^{\prime}}{M}\right\rceil\times\left\lceil\frac{W^{\prime}}{M}\right\rceil \times\left\lceil\frac{D^{\prime}}{M}\right\rceil$ regions at a given layer $l$ in the transformer encoder. Subsequently, in layer $l+1$, the partitioned window regions are shifted by $\left(\left\lfloor\frac{M}{2}\right\rfloor,\left\lfloor\frac{M}{2}\right\rfloor,\left\lfloor\frac{M}{2}\right\rfloor\right)$ voxels. In subsequent layers of $l$ and $l+1$ in the encoder, the outputs are calculated as
\begin{equation}
\begin{array}{l}
\hat{{z}}^{l}=\text{W-MSA}(\text{LN}({z}^{l-1}))+{z}^{l-1} \\
{z}^{l}=\text{MLP}(\text{LN}(\hat{{z}}^{l}))+\hat{{z}}^{l} \\
\hat{{z}}^{l+1}=\text{SW-MSA}(\text{LN}({z}^{l}))+{z}^{l} \\
{z}^{l+1}=\text{MLP}(\text{LN}(\hat{{z}}^{l+1}))+\hat{{z}}^{l+1}.
\end{array}
\label{eq:eq1}
\end{equation}
Here, $\text{W-MSA}$ and $\text{SW-MSA}$ are regular and window partitioning multi-head self-attention modules respectively;  $\hat{{z}}^{l}$ and $\hat{{z}}^{l+1}$ denote the outputs of $\text{W-MSA}$ and $\text{SW-MSA}$; $\text{MLP}$ and $\text{LN}$ denote layer normalization and Multi-Layer Perceptron respectively. For efficient computation of the shifted window mechanism, we leverage a 3D cyclic-shifting~\cite{liu2021swin} and compute self-attention according to
\begin{equation}
    \textnormal{Attention}(Q, K, V) = \textnormal{Softmax}\left(\frac{QK^{\top}}{\sqrt{d}}\right)V.
\label{eq:eq2}
\end{equation}
In which $Q, K, V$ denote queries, keys, and values respectively; $d$ represents the size of the query and key.

The Swin UNETR encoder has a patch size of $2 \times 2 \times 2$ and a feature dimension of $2\times2\times2\times4 =32$, taking into account the multi-modal MRI images with $4$ channels. The size of the embedding space $C$ is set to $48$ in our encoder. Furthermore, the Swin UNETR encoder has $4$ stages which comprise of $2$ transformer blocks at each stage. Hence, the total number of layers in the encoder is $L=8$. In stage $1$, a linear embedding layer is utilized to create $\frac{H}{2} \times \frac{W}{2} \times \frac{D}{2}$ 3D tokens. To maintain the hierarchical structure of the encoder, a patch merging layer is utilized to decrease the resolution of feature representations by a factor of $2$ at the end of each stage. In addition, a patch merging layer groups patches with resolution $2 \times 2 \times 2$ and concatenates them, resulting in a $4C$-dimensional feature embedding. The feature size of the  representations are subsequently reduced to $2C$ with a linear layer. Stage 2, stage 3 and stage 4, with resolutions of $\frac{H}{4} \times \frac{W}{4} \times \frac{D}{4}$, $\frac{H}{8} \times \frac{W}{8} \times \frac{D}{8}$ and $\frac{H}{16} \times \frac{W}{16} \times \frac{D}{16}$ respectively, follow the same network design. 

\subsection{Decoder}
Swin UNETR has a U-shaped network design in which the extracted feature representations of the encoder are used in the decoder via skip connections at each resolution. At each stage ${i}$ ($i \in \{0,1,2,3,4\})$ in the encoder and the bottleneck ($i=5$), the output feature representations are reshaped into size  $\frac{H}{2^{i}} \times \frac{W}{2^{i}} \times \frac{D}{2^{i}}$ and fed into a residual block comprising of two $3 \times 3 \times 3$ convolutional layers that are normalized by instance normalization~\cite{ulyanov2016instance} layers. Subsequently, the resolution of the feature maps are increased by a factor of $2$ using a deconvolutional layer and the outputs are concatenated with the outputs of the previous stage. The concatenated features are then fed into another residual block as previously described. The final segmentation outputs are computed by using a $1\times 1\times 1$ convolutional layer and a sigmoid activation function.

\begin{figure}[] 
 \centering
 \includegraphics[clip=true, trim=0pt 0pt 0pt 0pt, width=0.9\textwidth]{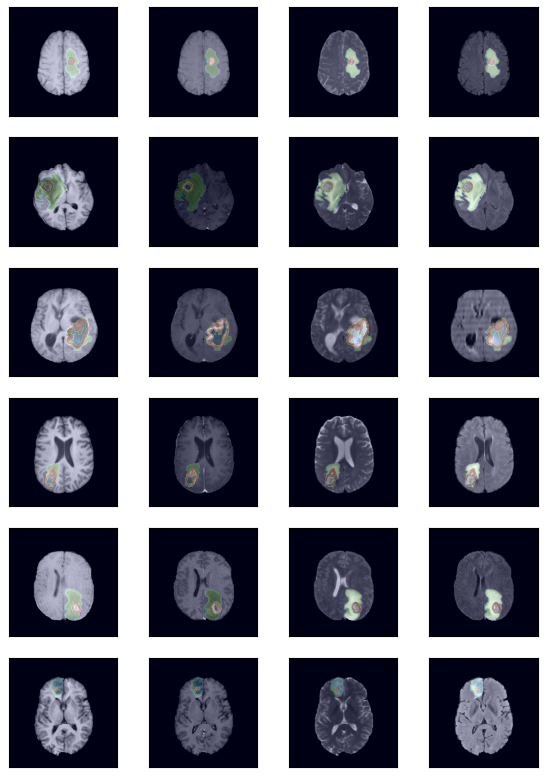}
 \caption{A typical segmentation example of the predicted labels whic are overlaid on T1, T1c, T2 and FLAIR MRI axial slices in each row. The first two rows depict $\sim${75}th percentile performance based on the Dice score. Rows 3 and 4 depict $\sim${50}th percentile performance while the last two rows are at $\sim${25}th percentile performance. The image intensities are on a gray color scale. The blue, red and green colors correspond to TC, ET and WT sub-regions respectively. Note that all samples have been selected from the BraTS 2021 validation set.}
 \label{fig:seg}
\end{figure}
\begin{table}[t]
 \centering
\resizebox{\columnwidth}{!}{
\begin{tabular}{lllllll}
\hline
Embed Dimension  & Feature Size & Number of Blocks  & Window Size & Number of Heads &  Parameters & FLOPs \\
\hline
768 & 48  & [2,2,2,2]   & [7,7,7]  & [3,6,12,24] & 61.98M & 394.84G   \\ 
\hline
\\
\end{tabular}%
}
\caption{Swin UNETR configurations.}
\label{tab:param}
\end{table}

\subsection{Loss Function}
We use the soft Dice loss function ~\cite{milletari2016v} which is computed in a voxel-wise manner as
\begin{equation}
\begin{split}
\mathcal{L}(G,Y) &= 1-\frac{2}{J}\sum_{j=1}^{J}\frac{\sum_{i=1}^{I} G_{i,j}Y_{i,j} }{\sum_{i=1}^{I}G^{2}_{i,j}+ \sum_{i=1}^{I}Y^{2}_{i,j}}.
\end{split}
\label{eq:ce}
\end{equation}
where $I$ denotes voxels numbers; $J$ is classes number; $Y_{i,j}$ and $G_{i,j}$ denote the probability of output and one-hot encoded ground truth for class $j$ at voxel $i$, respectively.

\subsection{Implementation Details}
Swin UNETR is implemented using PyTorch\footnote{\href{http://pytorch.org/}{http://pytorch.org/}} and MONAI\footnote{\href{https://monai.io/}{https://monai.io/}} and trained on a DGX-1 cluster with 8 NVIDIA V100 GPUs. Table~\ref{tab:param} details the configurations of Swin UNETR architecture, number of parameters and FLOPs. The learning rate is set to 0.0008. We normalize all input images to have zero mean and unit standard deviation according to non-zero voxels. Random patches of $128\times128\times128$ were cropped from 3D image volumes during training. We apply a random axis mirror flip with a probability of $0.5$ for all 3 axes. Additionally, we apply data augmentation transforms of random per channel intensity shift in the range $(-0.1,0.1)$, and random scale of intensity in the range $(0.9, 1.1)$ to input image channels. The batch size per GPU was set to 1. All models were trained for a total of 800 epochs with a linear warmup and using a cosine annealing learning rate scheduler. Fonr inference, we use a sliding window approach with an overlapping of 0.7 for neighboring voxels.

\subsection{Dataset and Model Ensembling}

The BraTS challenge aims to evaluate state-of-the-art methods for the semantic segmentation of brain tumors by providing a 3D MRI dataset with voxel-wise ground truth labels that are annotated by physicians~\cite{BratsAll2018,brats1,brats2,brats3,brats4}. The BraTS 2021 challenge training dataset includes 1251 subjects, each with four 3D MRI modalities: a) native (T1) and b) post-contrast T1-weighted (T1Gd), c) T2-weighted (T2), and d) T2 Fluid-attenuated Inversion Recovery (T2-FLAIR), which are rigidly aligned, and resampled to a $1\times1\times1$ mm isotropic resolution and skull-stripped. The input image size is $240\times240\times155$. The data were collected from multiple institutions using various MRI scanners. Annotations include three tumor sub-regions: the enhancing tumor, the peritumoral edema, and the necrotic and non-enhancing tumor core.  The annotations were combined into three nested sub-regions: Whole Tumor (WT), Tumor Core (TC), and Enhancing Tumor (ET). Fig.~\ref{fig:seg} illustrates typical segmentation outputs of all semantic classes. During this challenge, two additional datasets without the ground truth labels were provided for validation and testing phases. These datasets required participants to upload the segmentation masks to the organizers' server for evaluations. The validation dataset, which is designed for intermediate model evaluations, consists of 219 cases. Additional information regarding the testing dataset was not provided to participants.

Our models were trained on BraTS 2021 dataset with 1251 and 219 cases in the training and validation sets, respectively. Semantic segmentation labels corresponding to validation cases are not publicly available, and performance benchmarks were obtained by making submissions to the official server of BraTS 2021 challenge. We used five-fold cross-validation schemes with a ratio of 80:20. We did not use any additional data. The final result was obtained with an ensemble of 10 Swin UNETR models to improve the performance and achieve a better consensus for all predictions. The ensemble models were obtained from two separate five-fold cross-validation training runs.

\section{Results and Discussion}
 \label{sec:results}
We have compared the performance of Swin UNETR in our internal cross validation split against the winning methologies of previous years such as SegResNet~\cite{Myronenko18}, nnU-Net~\cite{Isensee2020nnUNetFB} and TransBTS~\cite{wang2021transbts}. The latter is a ViT-based approach which is tailored for the semantic segmentation of brain tumors. 

Evaluation results across all five folds are presented in Table~\ref{tab:five_fold}. The proposed Swin UNETR model outperforms all competing approaches across all 5 folds and on average for all semantic classes (e.g. ET, WT, TC). Specifically, Swin UNETR outperforms the closest competing approaches by $0.7\%, 0.6\%$ and $0.4\%$ for ET,WT and TC classes respectively and on average $0.5\%$ across all classes in all folds. The superior performance of Swin UNETR in comparison to other top performing models for brain tumor segmentation is mainly due to its capability of learning multi-scale contextual information in its hierarchical encoder via the self-attention modules and effective modeling of the long-range dependencies. 

Moreover, it is observed that nnU-Net and SegResNet have competitive benchmarks in these experiments, with nnU-Net demonstrating a slightly better performance. On the other hand, TransBTS, which is a ViT-based methodology, performs sub-optimally in comparison to other models. The sub-optimal performance of TransBTS could be attributed to its inefficient architecture in which the ViT is only utilized in the bottleneck as a standalone attention module, and without any connection to the decoder in different resolutions.

\begin{table}[t]
	\centering
	\caption{Five-fold cross-validation benchmarks in terms of mean Dice score values. ET, WT and TC denote Enhancing Tumor, Whole Tumor and Tumor Core respectively.}
	\label{tab:five_fold}
	\resizebox{\columnwidth}{!}{
	\begin{tabular}{l|c|c|c|c|c|c|c|c|c|c|c|c|c|c|c|c}
		\hline 
	& \multicolumn{4}{c|}{Swin UNETR} &
	\multicolumn{4}{c|}{nnU-Net} &
	\multicolumn{4}{c|}{SegResNet} &
	\multicolumn{4}{c}{TransBTS}\\ \hline
		Dice Score & ET & WT & TC& Avg. & ET & WT & TC & Avg.& ET & WT & TC & Avg.& ET & WT & TC & Avg. \\ \hline
		Fold 1 &  \bf{0.876} & \bf{0.929} & \bf{0.914}& \bf{0.906} & 0.866 & 0.921 & 0.902& 0.896&0.867 & 0.924 & 0.907& 0.899& 0.856 & 0.910 & 0.897& 0.883 \\
		Fold 2 & \bf{0.908} & \bf{0.938} & \bf{0.919}& \bf{0.921} & 0.899 & 0.933 & \bf{0.919}& 0.917& 0.900 & 0.933 & 0.915& 0.916& 0.885 & 0.919 &0.903& 0.902 \\
		Fold 3 & \bf{0.891} & \bf{0.931} & \bf{0.919}& \bf{0.913} & 0.886 & 0.929 & 0.914& 0.910& 0.884 & 0.927 & 0.917& 0.909 & 0.866 & 0.903 & 0.898& 0.889 \\
		Fold 4 & \bf{0.890} & \bf{0.937} & \bf{0.920}& \bf{0.915} & 0.886 & 0.927 & 0.914& 0.909& 0.888 & 0.921 & 0.916& 0.908& 0.868 & 0.910 & 0.901& 0.893  \\
		Fold 5 & \bf{0.891} & \bf{0.934} & \bf{0.917}& \bf{0.914} & 0.880 & 0.929 & \bf{0.917}& 0.909 & 0.878 & 0.930 & 0.912& 0.906& 0.867 & 0.915 & 0.893& 0.892 \\
		\hline
		Avg. & \bf{0.891}&	\bf{0.933}&	\bf{0.917}&	\bf{0.913}&	0.883&	0.927&	0.913&	0.908&	0.883&	0.927&	0.913&	0.907&	0.868&	0.911&	0.898&	0.891
 \\
		\hline
	\end{tabular}
	}
\end{table}

\begin{table}[t]
\centering
	\caption{BraTS 2021 validation dataset benchmarks in terms of mean Dice score and Hausdorff distance values. ET, WT and TC denote Enhancing Tumor, Whole Tumor and Tumor Core respectively.}
	\label{tab:val_set}
	\begin{tabular}{l|c|c|c|c|c|c}
		\hline
		& \multicolumn{3}{c|}{Dice} & \multicolumn{3}{c}{Hausdorff (mm)}  \\ \hline
		Validation dataset & ET & WT & TC & ET & WT & TC \\ \hline
		Swin UNETR & 0.858 & 0.926 & 0.885 & 6.016 & 5.831 & 3.770 \\
		\hline
	\end{tabular}
\end{table}

The segmentation performance of Swin UNETR in the BraTS 2021 validation set is presented in Table~\ref{tab:val_set}. According to the official challenge results\footnote{{\url{https://www.synapse.org/\#!Synapse:syn25829067/wiki/612712}}}, our benchmarks (Team: NVOptNet) are considered as one of the top-ranking methodologies across more than 2000 submissions during the validation phase, hence being the first transformer-based model to place competitively in BraTS challenges. In addition, the segmentation outputs of Swin UNETR for several cases in the validation set are illustrated in Fig.~\ref{fig:seg}. Consistent with quantitative benchmarks, the segmentation outputs are well-delineated for all three sub-regions.

Furthermore, the segmentation performance of Swin UNETR in the BraTS 2021 testing set is reported in Table~\ref{tab:test_set}. We observe that the segmentation performance of ET and WT are very similar to those of the validation benchmarks. However, the segmentation performance of TC is decreased by $0.9\%$.

\begin{table}[t]
\centering
	\caption{BraTS 2021 testing dataset benchmarks in terms of mean Dice score and Hausdorff distance values. ET, WT and TC denote Enhancing Tumor, Whole Tumor and Tumor Core respectively.}
	\label{tab:test_set}
	\begin{tabular}{l|c|c|c|c|c|c}
		\hline
		& \multicolumn{3}{c|}{Dice} & \multicolumn{3}{c}{Hausdorff (mm)}  \\ \hline
		Testing dataset & ET & WT & TC & ET & WT & TC \\ \hline
		Swin UNETR & 0.853 & 0.927 & 0.876 & 16.326 & 4.739 & 15.309 \\
		\hline
	\end{tabular}
\end{table}

\section{Conclusion}
 \label{sec:concl}
 In this paper, we introduced Swin UNETR which is a novel architecture for semantic segmentation of brain tumors using multi-modal MRI images. Our proposed model has a U-shaped network design and uses a Swin transformer as the encoder and CNN-based decoder that is connected to the encoder via skip connections at different resolutions. We have validated the effectiveness of our approach by in the BraTS 2021 challenge. Our model ranks among top-performing approaches in the validation phase and demonstrates competitive performance in the testing phase. We believe that Swin UNETR could be the foundation of a new class of transformer-based models with hierarchical encoders for the task of brain tumor segmentation.  

\bibliographystyle{splncs04}
\bibliography{ref}

\end{document}